\documentclass[12pt]{article}
\usepackage{epsfig}
\usepackage{amsfonts}
\usepackage{amsopn}
\usepackage{amsmath}
\usepackage{verbatim,amsthm}
\title
{
\vskip 20 pt
 On brane symmetries
}
\author{
 A. A. Zheltukhin 
  \thanks{E-mail: aaz@physto.se} 
  \\
Kharkov Institute of Physics and Technology, \\
1, Akademicheskaya St., Kharkov, 61108, Ukraine \\  
}

\date{}
\begin{document}

\maketitle
\begin{abstract}
The geometric approach to branes is reformulated in terms of 
gauge vector fields interacting with massless tensor multiplets 
in gravitational backgrounds.

\end{abstract}


Study of nonlinear dynamics of p-branes 
[1-14] \nocite{WHN, FI, WLN, BZ_0, hoppe2, Pol, CT, KY,HS, HN, AFP, hoppe6, AF, JU1} 
 as well as their quantization require new tools.
The geometric approach [15-17] \nocite{RL, Om, BN} originally 
developed for strings seems to be relevant for the problem.
 The gauge reformulation  \cite{Zgau} of this approach using  the 
ideas of Cartan \cite{Car}, Volkov \cite{Vol} and Faddeev \cite{SF} has shown 
 that strings in D-dim. spacetime form a closed sector of states of 
 the exactly integrable 
 two-dimensional $SO(1,1)\times SO(D-2)$ gauge model. 
  The geometric approach has turned out to be promising for investigation 
  of integrability of branes 
  PDEs [22-24] \nocite{TZ, ZT, Znpb}. 
Here we adopt the string gauge approach to p-branes and 
  constuct new gauge invariant models which have brane solutions.

\bigskip 

 1. A time-like $(p+1)$-dim. hypersurface $\Sigma_{p+1}$ embedded into the
 D-dim. Minkowski spacetime  with the signature $\eta_{mn}=(+,-,\ldots,-)$ 
 is described by its radius vector $\mathbf{x}(\xi^{\mu})$ parametrized by 
the coordinates $\xi^{\mu}=(\tau,\sigma^r), \, (r=1,2,..,p)$. 
Using a local orthonormal 
frame $\mathbf{n}_{A}(\xi^{\mu})=(\mathbf{n}_{i}, \mathbf{n}_{a})$ 
with  $A=(i,a)$, attached to $\Sigma_{p+1}$, one can expand the 
infinitesimal displacements 
 $d\mathbf{x}(\xi^{\mu})$ and 
$d\mathbf{n}_{A}(\xi^{\mu})$ in the local basis $\mathbf{n}_{A}(\xi^{\mu})$ at the point $\xi^{\mu}$
\begin{eqnarray}
d\mathbf{x}(\xi)=\omega^{i}(\xi)\mathbf{n}_{i}(\xi), \ \ \ \omega^{a}(\xi)=0,  \label{trl} \\
d\mathbf{n}_{A}(\xi)=-\omega_{A}^{\  B}(\xi)\mathbf{n}_{B}(\xi), \label{rot}
\end{eqnarray}
with the vectors 
$\mathbf{n}_{i}(\xi),\ (i,k=0,1,...,p)$ tangent and 
$\mathbf{n}_{a}(\xi)$, $(a,b=p+1,p+2,..., D-p-1)$ - normal to 
the hypersurface.
The choice $\omega^{a}=0$ of the normal displacement of $\mathbf{x}$ 
 breaks down the local Lorentz group $SO(1,D-1)$ of the moving frame 
to its  subgroup $SO(1,p)\times SO(D-p-1)$. 
Then the antisymmetric matrix differential form  $\omega_{AB}=- \omega_{BA}$ 
parametrized by $\xi^{\mu}$ and belonging to the Lie 
algebra of $SO(1,D-1)$ splits into three blocks 
\begin{eqnarray}\label{spl}
\omega_{A}{}^{B}\equiv\omega_{\mu A}{}^{B}d\xi^{\mu} 
=\left( \begin{array}{cc}
                       A_{\mu i}{}^{k}& W_{\mu  i}{}^{b} \\
                         W_{\mu a}{}^{ k} & B_{\mu a}{}^{b}
                              \end{array} \right)d\xi^{\mu} \ ,
\end{eqnarray}
where $A_{\mu i}{}^{k}$ and  $B_{\mu a}{}^{b}$ are transformed as the 
 gauge fields of the $SO(1,p)$ and $SO(D-p-1)$ groups on the base
  space $\Sigma_{p+1}$, respectively, and their field strengths   
 $F_{\mu\nu i}{}^{k}$ and  $H_{\mu\nu a}{}^{b}$ are
\begin{eqnarray}
F_{\mu\nu i}{}^{k}\equiv 
[D_{\mu}^{||},\,  D_{\nu}^{||}]_{i}{}^{k}=(\partial_{[\mu}A_{\nu]} + A_{[\mu}A_{\nu]})_{i}{}^{k}, 
 \label{F} \\
H_{\mu\nu a}{}^{b}\equiv 
[D_{\mu}^{\perp},\,  D_{\nu}^{\perp}]_{a}{}^{b}=
(\partial_{[\mu}B_{\nu]} + B_{[\mu}B_{\nu]})_{a}{}^{b}. 
 \label{H}
\end{eqnarray}
The derivative  $D_{\mu}^{||}$ in (\ref{F}) 
 is covariant with respect to the Lorentz gauge group 
$SO(1,p)$ of the subspaces tangent to $\Sigma_{p+1}$ 
\begin{eqnarray}\label{||der}
D_{\mu}^{||}\phi_{\nu}^{i}
=\partial_{\mu}\phi_{\nu}^{i}+ A_{\mu}{}^{i}{}_{k}\phi_{\nu}{}^{k}. 
\end{eqnarray}
The covariant derivative $D_{\mu}^{\perp}$ corresponds to the gauge 
group $SO(D-p-1)$ of rotations of the local subspaces orthogonal to $\Sigma_{p+1}$
\begin{eqnarray}\label{perpder}
D_{\mu}^{\perp}\phi_{\nu}^{a}
=\partial_{\mu}\phi_{\nu}^{a}+ B_{\mu}{}^{a}{}_{b}\phi_{\nu}{}^{b}.
\end{eqnarray}  

The off-diagonal blocks $W_{\mu  i}{}^{b}$ in (\ref{spl})
are transformed like charged vector multiplets of the gauge 
group $SO(1,p)\times SO(D-p-1)$ with their covariant derivatives 
\begin{eqnarray}\label{cd}
(D_{\mu} W_{\nu})_{i}{}^{a}= \partial_{\mu}W_{\nu i}{}^{a}+ A_{\mu i}{}^{k} W_{\nu k}{}^{a} + 
B_{\mu}{}^{a}{}_{b} W_{\nu i}{}^{b} 
\end{eqnarray}
including the gauge fields $A_{\mu i}{}^{k}$ and $B_{\mu a}{}^{b}$.

The integrability conditions of PDEs (\ref{trl}) and (\ref{rot}) are 
the Maurer-Cartan (M-C) equations
\begin{eqnarray}
d\wedge\omega_{A}+ \omega_{A}{}^{B}\wedge \omega_{B}=0, 
\label{intrl} \\
d\wedge\omega_{A}{}^{B} + \omega_{A}{}^{C}\wedge\omega_{C}{}^{B}=0  \label{inrot}
\end{eqnarray}
of the structure of the ambient D-dim. space with zero torsion and curvature, where the 
 symbols  $\wedge$ and $d\wedge$  mean the wedge product and external differential, respectively.

One can see that Eqs. (\ref{inrot}), called the Gauss-Codazzi (G-C) 
equations in the  differential geometry of surfaces, contain only the differential
 form $\omega_{A}{}^{B}$.
 The splitting (\ref{spl}) of the matrix indices $A \rightarrow (i,a)$ in (\ref{inrot})
results in the field representation of the G-C equations
\begin{eqnarray}
F_{\mu\nu i}{}^{k}= -(W_{[\mu} W_{\nu]})_{i}{}^{k},
\label{cF} \\
H_{\mu\nu a}{}^{b}= -(W_{[\mu} W_{\nu]})_{a}{}^{b},
\label{cH} \\
(D_{[\mu} W_{\nu]})_{i}{}^{a}=0 \label{ccd},
\end{eqnarray}
where $[\mu, \nu]$ means antisymmetrization in $\mu, \nu$, 
e.g. $\hat{W}_{[\mu}\hat{W}_{\nu]}\equiv \hat{W}_{\mu}\hat{W}_{\nu}-\hat{W}_{\nu}\hat{W}_{\mu}$.

For $p=1$ the above constraints coincide with the ones 
discussed upon the gauge reformulation of the geometric 
approach for strings \cite{Zgau}. This reformulation reveals 
an isomorphism between the Nambu-Goto string in D-dim. Minkowski space and 
the exactly solvable sector of the two-dim. 
 $SO(1,1)\times SO(D-2)$ gauge model including a massless scalar multiplet. 

Our main goal is to generalize the string case to p-branes which implies 
construction of a $(p+1)$-dim. $SO(1,p)\times SO(D-p-1)$ gauge model
 compatible with the G-C equations (\ref{cF}-\ref{ccd}). 
 This step does not suppose in advance any connections of such a model 
 with the existing models for p-branes, but only
  takes into account the independence of the constraints (\ref{cF}-\ref{ccd}) of
 the induced  metrics of hypersurfaces imbedded into flat spaces. 
 A class of new gauge actions compatible with the G-C constraints 
 is proposed in the next section.  

\bigskip 

2.  The desired $SO(1,p)\times SO(D-p-1)$ gauge-invariant
action has to describe the gauge and vector fields 
in an external gravitational field in $(p+1)$-dim. pseudo-Riemannian 
space with a metric $g_{\mu\nu}(\xi)$ parametrized by the
 coordinates $\xi^{\mu}$ (that will be later identified with the
 coordinates parametrizing the brane hypersurface $\Sigma_{p+1}$).
The metric $g_{\mu\nu}$ is not considered as a dynamical field
in contrast to the fields presented in the G-C constraints (\ref{cF}-\ref{ccd}).
The desired gauge and reparametrization invariant action has the form
\begin{eqnarray}\label{actn}
S= \gamma\int d^{p+1}\xi\sqrt{|g|}\, \, \mathcal{L}, \\
\mathcal{L}=
\frac{1}{4}Sp(F_{\mu\nu}F^{\mu\nu}) - \frac{1}{4}Sp(H_{\mu\nu}H^{\mu\nu}) \nonumber \\
+ \frac{1}{2}\hat{\nabla}_{\mu}W_{\nu}^{ia} \,  \hat{\nabla}^{\{\mu}W^{\nu\}}_{ia}
-\hat{\nabla}_{\mu}W^{\mu ia} \,  \hat{\nabla}_{\nu} W^{\nu}_{ia} + V,
\label{lagr}
\end{eqnarray}
where $\{\mu,\nu\}$ means symmetrization in $\mu$ and $\nu$, V 
 encodes nonlinear (self)interactions of the vector multiplet $W_{\mu}{}^{ia}$.  
The generalized covariant derivative $\hat{\nabla}_{\mu}$ in (\ref{lagr}) is 
\begin{eqnarray}\label{gcodr}
\hat{\nabla}_{\mu}W_{\nu ia}: =\partial_{\mu}W_{\nu ia} - \Gamma_{\mu\nu}^{\rho}W_{\rho ia} 
+ A_{\mu i}{}^{k}W_{\nu ka} + B_{\mu a}{}^{b}W_{\nu ib}
\end{eqnarray}
and extends the general covariant derivative including  only 
the Levi-Chivita connection
\begin{eqnarray}\label{cdrv}
\bigtriangledown_{\mu}W_{\nu ia}=\partial_{\mu}W_{\nu ia} - 
\Gamma_{\mu\nu}^{\rho}W_{\rho ia}, \ \ \ \ 
\bigtriangledown_{\mu}g_{\nu\rho}=0,
\end{eqnarray}
 where $\Gamma_{\mu\nu}^{\rho}=\Gamma_{\nu\mu}^{\rho}
=\frac{1}{2}g^{\rho\gamma}(\partial_{\mu}g_{\nu\gamma} + \partial_{\nu}g_{\mu\gamma}
- \partial_{\gamma}g_{\mu\nu})$ 
are the Cristoffel symbols.

The variation of $S$ (\ref{actn}) in the gauge and vector fields results in  
the following EOM
\begin{eqnarray}
\hat{\nabla}_{\mu} F^{\mu\nu}_{ik}
= -\hat{\nabla}_{\mu}(W^{[ \mu}_{ia}W^{\nu ] a}{}_{k}) 
- \frac{1}{2}W_{\mu  [i|a} \hat{\nabla}^{[\nu} W^{\mu]a}{}_{|k ]}, 
\label{maxF}   \\
\hat{\nabla}_{\mu} H^{\mu\nu}_{ab}
= -\hat{\nabla}_{\mu}(W^{[ \mu}_{ai}W^{\nu ] i}{}_{b}) 
- \frac{1}{2}W_{\mu  [a|i} \hat{\nabla}^{[\nu} W^{\mu]i}{}_{|b ]}
\label{maxH} \\ 
\hat{\nabla}_{\mu}\hat{\nabla}^{\{ \mu}W^{\nu \} ia}
=2\hat{\nabla}^{\nu}\hat{\nabla}_{\mu}W^{\mu ia}
+  \frac{\partial V}{\partial W_{\nu ia}}.
\label{eqgW}
\end{eqnarray}

With the help of shifted gauge field strengths  ${\cal F}^{\mu\nu}_{ik}$ 
and ${\cal H}^{\mu\nu}_{ab}$
\begin{eqnarray}
{\cal F}_{\mu\nu}^{ik}=(F_{\mu\nu} +W_{[\mu} W_{\nu]})^{ik}, 
\label{shftF} \\
{\cal H}_{\mu\nu}^{ab}=(H_{\mu\nu} +W_{[\mu} W_{\nu]})^{ab} 
\label{shftH}
\end{eqnarray}
one can present Eqs. (\ref{maxF}-\ref{eqgW}) in the compact form 
\begin{eqnarray}
\hat{\nabla}_{\mu} {\cal F}^{\mu\nu}_{ik}= 
- \frac{1}{2}W_{\mu  [i|a} \hat{\nabla}^{[\nu} W^{\mu]a}{}_{|k ]}, \ \ \ \ \ \ \ \
\label{maxF1}   \\
\hat{\nabla}_{\mu} {\cal H}^{\mu\nu}_{ab}= 
 - \frac{1}{2}W_{\mu  [a|i} \hat{\nabla}^{[\nu} W^{\mu]i}{}_{|b ]}, \ \ \ \ \ \ \ \
\label{maxH1}  \\
\hat{\nabla}_{\mu}\hat{\nabla}^{[\mu}W^{\nu] ia}
=- 2[\hat{\nabla}^{\mu} , \, \hat{\nabla}^{\nu}]
W_{\mu}^{ia}
+  \frac{\partial V}{\partial W_{\nu ia}}.
\label{eqgW1}
\end{eqnarray}
Further we take into account the  generalized first Bianchi identity 
\begin{eqnarray}\label{BI}
[\hat{\nabla}_{\mu} , \, \hat{\nabla}_{\nu}] 
=\hat{R}_{\mu\nu} + \hat{F}_{\mu\nu} + \hat{H}_{\mu\nu},
\end{eqnarray}
where the Riemann-Cristoffel 
tensor $\hat{R}_{\mu\nu}\equiv R_{\mu\nu}{}^{\gamma}{}_{\lambda}$ 
is defined  as
\begin{eqnarray}\label{Riem}
[\bigtriangledown_{\mu}, \, \bigtriangledown_{\nu}]V^{\gamma}
=R_{\mu\nu}{}^{\gamma}{}_{\lambda}V^{\lambda}
=: (\partial_{[\mu}\Gamma_{\nu]\lambda}^{\gamma} 
+\Gamma_{[\mu|\rho}^{\gamma}\Gamma_{|\nu] \lambda}^{\rho})V^{\lambda}.
\end{eqnarray}
The identity (\ref{BI}) allows to present Eq. (\ref{eqgW1}) 
 in the form  
\begin{eqnarray}\label{eqgW1s}                                    
\frac{1}{2}\hat{\nabla}_{\mu}\hat{\nabla}^{[\mu}W^{\nu] ia}
-  {\cal F}^{\mu\nu i}{}_{k} W_{\mu}^{ka} 
- {\cal H}^{\mu\nu a}{}_{b} W_{\mu}^{ib}   \nonumber \\
= \frac{1}{2} \frac{\partial V}{\partial W_{\nu ia}}
+  ([[W^{\mu}, W^{\nu}], W_{\mu}])^{ia} - R^{\mu\nu} W_{\mu}^{ia},
\end{eqnarray}
where $R_{\nu\lambda}:=R^{\mu}{}_{\nu}{}_{\mu\lambda}$ is the Ricci tensor. 
Using the relation 
\begin{eqnarray}\label{shftW}
\frac{1}{4}\frac{\partial}{\partial W_{\nu ia}}
(W_{\mu}[[W^{\mu}, W^{\rho}], W_{\rho}])^{i}{}_{i}
=([[W^{\mu}, W^{\nu}], W_{\mu}])^{ia} 
\end{eqnarray}
with the commutators of $\hat{W}_{\mu}$ in the r.h.s. we introduce 
 a shifted potential ${\cal V}$   
\begin{eqnarray}\label{shftV}
{\cal V}= V + \frac{1}{2}Sp(W_{\mu}[[W^{\mu}, W^{\rho}], W_{\rho}]),
\end{eqnarray}
where the trace $Sp(W_{\mu}[[W^{\mu}, W^{\rho}], W_{\rho}]):
= (W_{\mu}[[W^{\mu}, W^{\rho}], W_{\rho}])^{i}{}_{i}$.

As a  result,  EOM (\ref{maxF1}), (\ref{maxH1}) and (\ref{eqgW1s})
 take the following form
\begin{eqnarray}
\hat{\nabla}_{\mu} {\cal F}^{\mu\nu}_{ik}= 
- \frac{1}{2}W_{\mu  [i|a} \hat{\nabla}^{[\nu} W^{\mu]a}{}_{|k ]}, \ \ \ \ \ \ \ \ \ \ \ \ \ \ \ \ 
\label{maxF1'} \\
\hat{\nabla}_{\mu} {\cal H}^{\mu\nu}_{ab}= 
 - \frac{1}{2}W_{\mu  [a|i} \hat{\nabla}^{[\nu} W^{\mu]i}{}_{|b ]}, \ \ \ \ \ \ \ \ \ \ \ \ \ \ \ \ 
\label{maxH1'} \\
\frac{1}{2}\hat{\nabla}_{\mu}\hat{\nabla}^{[\mu}W^{\nu] ia}
+  {\cal F}^{\mu\nu i}{}_{k} W_{\mu}^{ka} 
+ {\cal H}^{\mu\nu a}{}_{b} W_{\mu}^{ib}   
= \frac{1}{2} \frac{\partial {\cal V}}{\partial W_{\nu ia}}
- R^{\mu\nu} W_{\mu}^{ia} .
\label{eqgW1ss}
\end{eqnarray}

Then we observe that the first-order PDEs which coincide with (\ref{cF}-\ref{ccd})
\begin{eqnarray}\label{FHWsol} 
{\cal F}_{\mu\nu}^{ik}=0,\ \ \ \ \ {\cal H}_{\mu\nu}^{ab}=0, 
\ \ \ \ \ \hat{\nabla}^{[\mu}W^{\nu]}_{ia}=0,
\end{eqnarray}
form a particular solution of Eqs. (\ref{maxF1'}-\ref{eqgW1ss})
on condition that
\begin{eqnarray}\label{eqR}
\frac{1}{2}\frac{\partial{\cal V}}{\partial W_{\nu ia}}
- R^{\mu\nu} W_{\mu}^{ia}=0.
\end{eqnarray}
 Due to independence of the  Ricci tensor $ R^{\mu\nu}$  
of $W_{\nu ia}$,   Eq. (\ref{eqR})
 allows to restore ${\cal V}$ 
\begin{eqnarray}\label{VRsolu}
{\cal V}=  R^{\mu\nu} W_{\mu}^{ia}W_{\nu ia}.
\end{eqnarray}

Thus, we find that the action (\ref{actn}) with the  Lagrangian density 
\begin{eqnarray} 
\mathcal{L}=
\frac{1}{4}Sp(F_{\mu\nu}F^{\mu\nu}) - \frac{1}{4}Sp(H_{\mu\nu}H^{\mu\nu}) \nonumber \\
+ \frac{1}{2}\hat{\nabla}_{\mu}W_{\nu}^{ia} \,  \hat{\nabla}^{\{\mu}W^{\nu\}}_{ia}
-\hat{\nabla}_{\mu}W^{\mu ia} \,  \hat{\nabla}_{\nu} W^{\nu}_{ia} 
\label{lagrR}  \\
+ R^{\mu\nu} W_{\mu}^{ia}W_{\nu ia}
- \frac{1}{2}Sp(W_{\mu}[[W^{\mu}, W^{\rho}], W_{\rho}])
\nonumber
\end{eqnarray}
yields the nonlinear Euler-Lagrange equations 
\begin{eqnarray}
\hat{\nabla}_{\mu} {\cal F}^{\mu\nu}_{ik}= 
- \frac{1}{2}W_{\mu  [i|a} \hat{\nabla}^{[\nu} W^{\mu]a}{}_{|k ]}, 
\label{maxF12} \\
\hat{\nabla}_{\mu} {\cal H}^{\mu\nu}_{ab}= 
 - \frac{1}{2}W_{\mu  [a|i} \hat{\nabla}^{[\nu} W^{\mu]i}{}_{|b ]},
\label{maxH12} \\
\frac{1}{2}\hat{\nabla}_{\mu}\hat{\nabla}^{[\mu}W^{\nu] ia}
+  {\cal F}^{\mu\nu i}{}_{k} W_{\mu}^{ka} 
+ {\cal H}^{\mu\nu a}{}_{b} W_{\mu}^{ib}   
=0
\label{eqgW1ss2}
\end{eqnarray}
for the gauge  $A_{\mu i}{}^{k}, \,   
B_{\mu a}{}^{b}$ and vector $W_{\mu ia}$ fields in a given external 
gravitational field  $g_{\mu\nu}(\xi^{\rho})$. 

It is  easy to see that 
Eqs. (\ref{maxF12}-\ref{eqgW1ss2})  have the particular solution 
(\ref{FHWsol}) which coincides with the G-C
constraints  (\ref{cF}), (\ref{cH}) and (\ref{ccd}). 

This solves the stated problem of the construction of gauge invariant model 
compatible with embedded hypesurfaces using the Gauss mapping.
In addition note that the action (\ref{actn}) with $\mathcal{L}$ 
(\ref{lagrR}) looks like a natural generalization of the four-dim. Dirac  scale-invariant 
gravity theory with the dynamical dilaton and gravitational field $g_{\mu\nu}$ (see e.g. \cite{PeP}). 

The above-said hints at consideration of the
 $(p+1)$-dim. spacetime of the gauge model defined by (\ref{actn}), (\ref{lagrR})
  as a (p+1)-dim. world hypersurface swept by a p-brane in D-dim. Minkowski space. 
Our next step is to prove that the conjecture follows 
from the remaining Maurer-Cartan eqs. (\ref{intrl}) and to find 
 the corresponding modification of the proposed model.

\bigskip

3.  To prove the mentioned statement we come back to 
the M-C Eqs. (\ref{intrl}) and split their 
matrix indices $A \rightarrow (i,a)$. 
This yields  the following equations
\begin{eqnarray}
D_{[\mu}^{||}\omega_{\nu]}^{i}=0,  \label{csym} \\
\omega_{[\mu}^{i}W_{\nu]ia}=0 \label{lcntr}
\end{eqnarray}
with the derivative $D_{\mu}^{||}$ defined by (\ref{||der}).
As shows $d\mathbf{x}$ squaring, the object $\omega_{\mu}^{i}$ 
 plays the role of a $(p+1)$-bein for the hypersurface $\Sigma_{p+1}$
which connects its  orthonormal frame $\mathbf{n}_{i}$ with 
the  local natural frame $\mathbf{e}_{\mu}$, and represents 
the metric $G_{\mu\nu}(\xi^\rho)$ 
of $\Sigma_{p+1}$  by the quadratic form 
\begin{eqnarray}\label{metr}
\omega_{\mu}^{i}\omega^{\mu}_{k}=\delta^{i}_{k}, \ \ \
\mathbf{e}_{\mu}=\omega_{\mu}^{i}\mathbf{n}_{i}, \ \ \ 
G_{\mu\nu}=\omega_{\mu}^{i}\eta_{ik}\omega_{\nu}^{k}.
\end{eqnarray}

One can solve the constraints (\ref{lcntr}) and 
 express $W_{\mu  i}{}^{a}$ 
 in terms of the symmetric components $l_{\mu\nu}{}^{a}$ 
 of the second fundamental form of $\Sigma_{p+1}$
\begin{eqnarray}\label{2frm}
W_{\mu i}{}^{a}= -l_{\mu\nu}{}^{a}\omega^{\nu}_{i}, \ \ \ \
l_{\mu\nu}{}^{a}:=\mathbf{n}^{a}\partial_{\mu\nu}\mathbf{x}.
\end{eqnarray}

The general solution of the constraints (\ref{csym}) is equivalent 
 to the "tetrade postulate"
\begin{eqnarray}\label{tetpo}
\nabla_{\mu}^{||}\omega_{\nu}^{i}
\equiv\partial_{\mu}\omega_{\nu}^{i} - \Gamma_{\mu\nu}^{\rho}\omega_{\rho}^{i}
 + A_{\mu}{}^{i}{}_{k}\omega_{\nu}^{k} = 0 
\end{eqnarray}
which identifies the gauge connection $A_{\mu}{}^{i}{}_{k}$ with 
the background metric connection $\Gamma_{\mu\nu}^{\rho}$ 
by means of the gauge transformation 
\begin{eqnarray}\label{gtA}
\Gamma_{\nu\lambda}^{\rho}
=\omega_{i}^{\rho}A_{\nu}{}^{i}{}_{k}\omega^{k}_{\lambda}
+\partial_{\nu}\omega^{k}_{\lambda}\omega^{\rho}_{k}
\equiv\omega^{\rho}_{i}D_{\nu}^{||}\omega^{i}_{\lambda}.
\end{eqnarray}
Therefore, the hypersurface metric $G_{\mu\nu}$ has to be identified 
with the backgroud metric $g_{\mu\nu}$ introduced $\it {ad \, hock}$ in 
the gauge invariant action (\ref{actn}). 

Then the Riemann tensor $R_{\mu\nu}{}^{\gamma}{}_{\lambda}$ (\ref{Riem})  
and the field strength $F_{\mu\nu i}{}^{k}$ (\ref{F}) become dependent 
\begin{eqnarray}\label{FRcon}
R_{\mu\nu}{}^{\gamma}{}_{\lambda}
= \omega^{\gamma}_{i}F_{\mu\nu}{}^{i}{}_{k}\omega^{k}_{\lambda}, \ \ \ \ 
R_{\nu\lambda}
=  \omega^{\mu}_{i}F_{\mu\nu}{}^{i}{}_{k}\omega^{k}_{\lambda},
\end{eqnarray}
and the use of the G-C constraint (\ref{cF}) for $F_{\mu\nu}{}^{i}{}_{k}$ 
 allows to express the Ricci tensor as 
\begin{eqnarray}\label{Ric}
R^{\nu\lambda}
= - \omega_{\mu}^{i}(W^{[\mu}W^{\nu]})_{i}{}^{k}\omega_{k}^{\lambda}.
\end{eqnarray}  
Taking into account (\ref{metr}-\ref{FRcon}) permits to 
transit from the gauge  $A_{\nu ik}$ and vector $W_{\mu i}{}^{a}$ 
fields to the Cristoffel symbols 
and $l_{\mu\nu}{}^{a}= -\omega_{\nu}^{i}W_{\mu i}{}^{a}$, respectively,
 that  transforms (\ref{cF}-\ref{ccd}) into 
\begin{eqnarray}
R_{\mu\nu}{}^{\gamma}{}_{\lambda}=l_{[\mu}{}^{\gamma a} l_{\nu]\lambda a},
\label{cRl} \\
H_{\mu\nu }{}^{ab}= l_{[\mu}{}^{\gamma a} l_{\nu]\gamma}{}^{b},
\label{cH2} \\ 
\nabla_{[\mu}^{\perp}l_{\nu]\rho a}=0, \ \ \ \ \ \
\label{ccd'}
\end{eqnarray}
where 
$ 
\nabla_{\mu}^{\perp}l_{\nu\rho}{}^{a}:= \partial_{\mu}l_{\nu\rho}{}^{a}
- \Gamma_{\mu\nu}^{\lambda} l_{\lambda\rho}{}^{a} 
-\Gamma_{\mu\rho}^{\lambda} l_{\nu\lambda}{}^{a} + B_{\mu}^{ab}l_{\nu\rho b}.
$  

As is seen, exclusion of $F_{\mu\nu i}{}^{k}$ 
 transforms the constraint (\ref{cF}) into (\ref{cRl}) which generalizes 
the $\it {Gauss \, Theorema  \, Egregium}$ for a (p+1)-dim. hypersurface 
embedded into the D-dim. Minkowski space. 
The absence of $F_{\mu\nu i}{}^{k}$ allows not to consider 
 the group $SO(1,p)$ as an explicit symmetry of the desired action. 
 As a result, we obtain the following  $SO(D-p-1)$ gauge-invariant action 
 in a gravitational background possessing the solution (\ref{cRl}-\ref{ccd'}) 
 \begin{eqnarray}
S= \gamma\int d^{p+1}\xi\sqrt{|g|} \, \, \mathcal{L} ,  \nonumber 
  \ \  \ \  \ \  \ \  \ \  \ \  \ \   \ \  \ \  \ \  \ \ \ \  \ \  \ \ 
\\
\mathcal{L}= 
- \frac{1}{4}Sp(H_{\mu\nu}H^{\mu\nu}) 
+ \frac{1}{2}\nabla_{\mu}^{\perp}l_{\nu\rho a}\nabla^{\perp \{\mu}l^{\nu\}\rho a}
-\nabla_{\mu}^{\perp}l^{\mu}_{\rho a}\nabla_{\nu}^{\perp}l^{\nu\rho a}   
\nonumber 
\\
- \frac{1}{2} Sp(l_{a}l_{b}) Sp(l^{a}l^{b})
+ Sp(l_{a}l_{b}l^{a}l^{b})
- Sp(l_{a}l^{a}l_{b}l^{b}). \ \  \ \  \ \  \ \  \ \  \ \  \  \ \  \ \ 
\label{actnl} 
\end{eqnarray}

To prove this let us consider the following action
\begin{eqnarray}
S= \gamma\int d^{p+1}\xi\sqrt{|g|} \{
- \frac{1}{4}Sp(H_{\mu\nu}H^{\mu\nu})\nonumber 
\\
+ \frac{1}{2}\nabla_{\mu}^{\perp}l_{\nu\rho a}\nabla^{\perp \{\mu}l^{\nu\}\rho a}
-\nabla_{\mu}^{\perp}l^{\mu}_{\rho a}\nabla^{\perp}_{\nu}l^{\nu\rho a}  + V \},
\label{actn2}
\end{eqnarray}
Variation of (\ref{actn2}) in the dynamical 
fields $l_{\mu\nu}{}^{a}, \, B_{\mu}{}^{ab}$ gives their EOM
\begin{eqnarray}
\nabla^{\perp}_{\nu} {\cal H}^{\nu\mu}_{ab}= 
 \frac{1}{2}l_{\nu\rho[a}\nabla^{\perp[\mu} l^{\nu]\rho}{}_{b ]}, \ \ \ \ \ \ \ \ \ \ \ \ \ \
\label{maxH2'} \\
\frac{1}{2}\nabla^{\perp}_{\mu}\nabla^{\perp[\mu}l^{\nu]\rho a}
=-[\nabla^{\perp\mu},  \nabla^{\perp\nu}] l_{\mu}{}^{\rho a}
+\frac{1}{2} \frac{\partial {V}}{\partial l_{\nu\rho a}},   
\label{eqgW2ss}
\end{eqnarray}
where ${\cal H}_{\mu\nu}^{ab}:= H_{\mu\nu}^{ab} - l_{[\mu}{}^{\gamma a} l_{\nu]\gamma}{}^{b}$. 
 Equations (\ref{maxH2'}-\ref{eqgW2ss}) have the G-C constraints (\ref{cH2}-\ref{ccd'}) 
\begin{eqnarray}\label{Hlsol} 
{\cal H}_{\mu\nu}^{ab}=0, \ \ \ \ \  
\nabla_{[\mu}^{\perp}l_{\nu]\rho a}=0, 
\end{eqnarray}
 as their particular solution provided that
\begin{eqnarray}\label{eqV} 
\frac{1}{2} \frac{\partial {V}}{\partial l_{\nu\rho a}}
=[\nabla^{\perp\mu},  \nabla^{\perp\nu}] l_{\mu}{}^{\rho a}.
\end{eqnarray}

With the help of G-C  Eqs. (\ref{cRl}-\ref{ccd'}) and the Bianchi identity 
\begin{eqnarray}\label{BIl}
[\nabla^{\perp}_{\gamma} , \, \nabla^{\perp}_{\nu}] l^{\mu\rho a}
=R_{\gamma\nu}{}^{\mu}{}_{\lambda} l^{\lambda\rho a}  
+ R_{\gamma\nu}{}^{\rho}{}_{\lambda} l^{\mu\lambda a} 
+H_{\gamma\nu}{}^{a}{}_{b} l^{\mu\rho b}  
\end{eqnarray}
one can transform (\ref{eqV}) into solvable equation for the 
 self-interaction potential $V$ 
\begin{eqnarray}\label{eqVl} 
\frac{1}{2} \frac{\partial {V}}{\partial l_{\nu\rho a}}
=(l^{a}l^{b})^{\rho\nu}Sp(l_{b}) 
+ (2l_{b}l^{a}l^{b}-l^{a}l_{b}l^{b}-l_{b}l^{b}l^{a})^{\rho\nu}
-l^{\rho\nu b}Sp(l_{b}l^{a}). 
\end{eqnarray} 

Equation (\ref{eqVl}) has the following solution for V 
accompanied by the trace constraints 
\begin{eqnarray}\label{solVl} 
V=- \frac{1}{2} Sp(l_{a}l_{b}) Sp(l^{a}l^{b})
+ Sp(l_{a}l_{b}l^{a}l^{b}) - Sp(l_{a}l^{a}l_{b}l^{b}), \ \ \   
Sp(l_{a})=0.
\end{eqnarray} 

The constraints $Sp(l_{a})=0$ express the well-known algebraic conditions 
of minimality for a $(p+1)$-dim. hypersurface embedded 
into the Minkowski spaces. These conditions are equivalent to the nonlinear 
equations of motion of p-branes
\begin{equation}\label{Box}
\Box^{(p+1)}\mathbf{x}=0, 
\end{equation}
where  $\Box^{(p+1)}:
=\frac{1}{\sqrt{|G|}}\partial_{\alpha} \sqrt{|G|}G^{\alpha\beta}\partial_{\beta}$ 
is the reparametrization invariant Laplace-Beltrami operator on  $\Sigma_{p+1}$ \cite{Znpb}.

Eq. (\ref{Box}) follows from the Dirac action for p-branes with
 the minimal hypersurfaces in the Minkowski spacetime
\begin{equation}\label{Dirbr}
S=T\int d^{p+1}\xi \sqrt{|G|},
\end{equation}
where $G$ is the determinant of the induced metric
$G_{\alpha \beta}:=\partial_{\alpha}\mathbf{x}\partial_{\beta}\mathbf{x}$. 

It proves that the $SO(D-p-1)$ gauge-invariant action (\ref{actnl}) 
for the interacting gauge and tensor fields $B_{\mu}^{ab}$ and $l^{a}_{\mu\nu}$, 
respectively, in a gravitational background has the particular solution 
 presented by the first-order Gauss-Codazzi PDEs (\ref{cRl}-\ref{ccd'}). 
 The solution describes minimal $(p+1)$-dim. hypersurfaces embedded 
 into D-dim. Minkowski spacetime.

\medskip  
{\bf Summary}: 
The gauge reformulation of the geometric approach to $(p+1)$-dimensional 
hypersurfaces embedded into D-dimensional Minkowski space was proposed. 
The new set of $SO(1,p)\times SO(D-p-1)$-invariant gauge 
models possessing exact solutions for gauge fields and vector 
multiplets in gravitational backgrounds, was constructed.
The Dirac p-branes were shown to be the solutions of 
$(p+1)$-dimensional gauge model presented by the Gauss-Codazzi constraints 
for the $SO(D-p-1)$ gauge vector fields and massless tensor multiplets 
in curved backgrounds.

\bigskip 

\noindent{\bf Acknowledgments}

I am grateful to E. Ivanov and  V. Pervushin 
for interesting remarks, A. Rosly for valuable comments and 
sending the paper \cite{Hit}, to Physics Department of Stockholm University and 
 Nordic Institute for Theoretical Physics NORDITA for kind hospitality and support.


\begin{thebibliography}{99} 

\bibitem{FI}
E. Floratos and  J. Illipoulos, \textit{A note on the classical symmetries
 of the closed bosonic membranes}, Phys. Lett. B 201, 237 (1988).

\bibitem{WHN}
B. de Witt, J. Hoppe and G. Nicolai, \textit{On the quantum mechanics of supermembranes} 
Nucl.  Phys. B, [FS23] 305, 545 (1988).

\bibitem{WLN}
B. de Witt, M. Lusher and  G. Nicolai, \textit{The supermembrane is unstable}
Nucl.  Phys. B 320, 135 (1989). 

\bibitem{BZ_0}
I.A. Bandos and A.A. Zheltukhin, 
 \textit{Null super p-branes quantum theory in four-dimensional space-time} 
   Fortschr. Phys.  41, 619 (1993); 
 \\  \textit{N=1 super p-branes in twistor-like Lorentz harmonic formulation} 
Class. Quant. Grav.  12, 609 (1995). 

\bibitem{hoppe2} 
M. Bordemann, J. Hoppe, 
\textit{The Dynamics of Relativistic
Membranes I: Reduction to 2-dimensional Fluid Dynamics},  
  Phys. Lett. B 317, 315 (1993); 
\textit{The Dynamics of Relativistic
Membranes II: Nonlinear Waves and Covariantly Reduced Membrane Equations},
 ibid. B 325, 359 (1994).

\bibitem{Pol} 
J. Polchinski, \textit{Dirichlet-branes and Ramond-Ramond charges},
 Phys. Rev. Lett. 75, 4724 (1995). 

\bibitem{CT}
P.A. Collins and R. W. Tucker,  \textit{Transversity of a massless relativistic membrane}, 
Nucl. Phys. B 112, 150 (1976).

\bibitem{KY} 
K. Kikkawa and M. Yamasaki,  \textit{Can the membrane be a unification model?}, 
Progr. Theor. Phys. 76, 1379 (1986).

\bibitem{HS} 
P. S. Howe and E. Sezgin, 
 \textit{Superbranes},
Phys. Lett. B 390, 441 (1997). 

\bibitem{HN}
J. Hoppe and H. Nicolai, \textit{Relativistic Minimal Surfaces},
Phys. Lett. B. 196, 451 (1987). 

\bibitem{AFP}
M. Axenidis, E.G. Floratos and L. Perivolaropoulos, 
\textit{Rotating toroidal branes in supermembrane and matrix theory},
  Phys. Rev. D 66, 085006 (2002). 

\bibitem{hoppe6}
J. Arnlind, J. Hoppe and  S. Theisen,
\textit{Spinning membranes}, Phys. Lett. B 599, 118 (2004). 
 
\bibitem{AF}
 M.~Axenides and E.~G.~Floratos,
 \textit{Euler top dynamics of Nambu-Goto p-branes}, 
  JHEP  0703, 093 (2007).
 
 \bibitem{JU1}
J. Hoppe,
 \textit{U(1)-invariant membranes and singularity formation},
Compl. Anal. Oper. Theory 3, 419 (2009).
 
 \bibitem{RL}
F. Lund and T. Regge, \textit{Unified Approach to Strings and Vortices with Soliton Solution},   
Phys. Rev. D14, 1524 (1976). 

\bibitem{Om}
R. Omnes,    
\textit{A New Geometric Approach to the Relativistic String},   
Nucl. Phys. B 149, 269 (1979). 

\bibitem{BN}
B.M. Barbashov, V.V. Nesterenko,
\textit{Differential geometry and nonlinear field models},
 Fortschr. Phys. 28, 427 (1980) ; 
\\
\textit{Introduction to the Relativistic String Theory} 
(World Scientific Pub. Co. Inc., 1990).

\bibitem{Zgau}
A.A. Zheltukhin, 
\textit{Connection between a relativistic string and two-dimensional field model}, 
 Sov. J. Nucl. Phys. 33, 927 (1981);  
 \textit{Classical relativistic string as an exactly solvable sector 
 of $SO(1,1)\times SO(2)$ gauge model}, 
 Phys. Lett. B 116, 147 (1982); 
\textit{Classical relativistic string as a two-dimensional $SO(1,1)\times SO(2)$ gauge model}
, Theor. Math. Phys. 52, 666 (1982); 
\textit{Gauge description and nonlinear string equations in D-dimensional space-time}, 
 ibid.  56, 785 (1984).

\bibitem{Car}
E. Cartan,  \textit{Riemannian Geometry in an Orthogonal Frame}
  (World Scientific, Singapore, 2001).

\bibitem{Vol}
D.V. Volkov, Phys. of Element. Particles and Atomic Nuclei. 4, 3 (1973).
\bibitem{SF}
M.A. Semenov-Tyan-Shansky and L.D. Faddeev,  
\textit{On theory of nonlinear chiral fields}, 
Vestnik  St. Petersburg Univ. 13(3), (1977) [in Russian].

\bibitem{TZ}
M. Trzetrzelewski and A.A. Zheltukhin, 
 \textit{Exact solutions for U(1) globally invariant membranes}, 
 Phys. Lett. B 679, 523 (2009).

\bibitem{ZT}
A.A. Zheltukhin and M. Trzetrzelewski,  
\textit{U(1)-invariant membranes: the geometric formulation,
Abel and pendulum differential equations},
 J. Math. Phys.  51, 062303 (2010).

\bibitem{Znpb} 
A.A. Zheltukhin, \textit{Toroidal p-branes, anharmonic oscillators 
and (hyper)elliptic solutions}, Nucl. Phys. B 858, 142 (2012); 
\textit{Laplace-Beltrami operator and exact solutions for branes},
ibid. B 867,  763 (2013);
   \textit{Generalized Hooke law for relativistic membranes and p-branes},  
Prob. Atomic Sci. Technol.  1, 7 (2012);
\textit{On nonlinearity of $p$-brane dynamics},
Int. J. Geom. Meth. Mod. Phys. 9, 1261017 (2012).

\bibitem{Dir} 
P.A.M. Dirac,  \textit{Long Range Forces and Broken Symmetries},
Proc. R. Soc. Lond. A 333. 403 (1973).

\bibitem{PeP} 
V. Pervushin, A. Pavlov. 
\textit{Principles of Quantum Universe} Saarbrucken 
(LAP LAMBERT Academic Publishing, Germany, 2013). 

\bibitem{Hit}
N. Hitchin.  \textit{The self-duality equations on 
a Riemann surface}//
Proc. London Math. Soc. 55(3) 59 (1987).

\end{thebibliography}
\end{document}